\documentclass[twocolumn]{L98-59b}
\usepackage{appendix}
\usepackage{threeparttable}
\usepackage{multirow}
\usepackage{color}
\usepackage{bm}
\usepackage{natbib}

\shorttitle{Hubble WFC3 Spectroscopy of the Rocky Planet L~98-59~b}
\shortauthors{Zhou et al.}

\graphicspath{{./}{figures/}}

\begin{document}

\title{Hubble WFC3 Spectroscopy of the Rocky Planet L~98-59~b: No Evidence for a Cloud-Free Primordial Atmosphere}

\author{Li Zhou }
\affiliation{School of Physics and Astronomy, Sun Yat-sen University, Zhuhai 519082, China}
\affiliation{Center of CSST in the great bay area, Sun Yat-sen University, Zhuhai 519082, China}

\author{Bo Ma}
\email{mabo8@mail.sysu.edu.cn}
\affiliation{School of Physics and Astronomy, Sun Yat-sen University, Zhuhai 519082, China}
\affiliation{Center of CSST in the great bay area, Sun Yat-sen University, Zhuhai 519082, China}

\author{Yonghao Wang}
\affiliation{School of Physics and Astronomy, Sun Yat-sen University, Zhuhai 519082, China}
\affiliation{Center of CSST in the great bay area, Sun Yat-sen University, Zhuhai 519082, China}

\author{Yinan Zhu}
\affiliation{National Astronomical Observatories, Chinese Academy of Sciences, Beijing 100101, China}

\begin{abstract}
We are using archived data from HST of transiting exoplanet L~98-59~b to place constraints on its potentially hot atmosphere. 
We analyze the data from five transit visits and extract the final combined transmission spectrum using Iraclis. 
Then we use the inverse atmospheric retrieval code TauREx to analyze the combined transmission spectrum. 
There is a weak absorption feature near 1.40~$\mu m$ and 1.55~$\mu m$ in the transmission spectrum, which can be modeled by a cloudy atmosphere with abundant HCN. 
However, the unrealistically high abundance of HCN derived cannot be explained by any equilibrium chemical model with reasonable assumptions.
Thus, the likeliest scenario is that L~98-59~b has a flat, featureless transmission spectrum in the WFC3/G141 bandpass due to a thin atmosphere with high mean molecular weight, an atmosphere with an opaque aerosol layer, or no atmosphere, and it is very unlikely for L~98-59~b to have a clear hydrogen-dominated primary atmosphere.
Due to the narrow wavelength coverage and low spectral resolution of HST/WFC3 G141 grism observation, we cannot tell these different scenarios apart. 
Our simulation shows future higher precision measurements over wider wavelengths from the James Webb Space Telescope (JWST) can be used to better characterize the planetary atmosphere of L~98-59~b.
\end{abstract}

\section{Introduction} \label{sec:intro}
With the development of instruments and observing techniques, exoplanet science is one of the fastest developing sciences in astronomy. As more and more exoplanets are discovered, studying the detailed properties of individual exoplanets is just as important and popular as discovering more exoplanets.  
For example, characterization of an exoplanet's atmosphere is vital for understanding the interior and global properties of a planet. 
There are more and more studies about the atmospheres of gaseous giant planets. However, due to the lower transit depth, the atmospheres of terrestrial planets are more difficult to detect using transmission spectra than those of gas giant planets. Thus,  atmospheric studies of small rocky planets orbiting nearby bright stars are still pretty rare. 

The primary atmosphere of a terrestrial planet may have been removed or replaced by a secondary atmosphere with a higher mean molecular weight \citep{2019ARA&A..57..617M}, which makes it much harder to trace where the planet was born.
During the process of establishing a secondary atmosphere, a variety of related physical and chemical processes can provide us with the opportunity to detect bio-signatures on possible habitable planets. 

For now, there are only a few tentative studies about the  atmospheres of small planets using spectral data from the Hubble Space Telescope (HST)/Wide Field Camera 3 (WFC3). Some of these studies are summarized as below.
The highly irradiated super-Earth, 55 Cancri e, was found to have hydrogen cyanide (HCN, \citealt{2016ApJ...820...99T}) in its atmosphere.
TRAPPIST-1 b and c, two Earth-sized terrestrial planets discovered in 
\citet{2016Natur.537...69D}, do not have a cloud-free hydrogen-dominated atmosphere.
There are no obvious absorption features in the transmission spectra of GJ 1214b ($\rm T_{eq}$ = 596 K, R = 2.742 $\rm R_\oplus$, \citealt{2021arXiv210714732C},) and HD 97658b ($\rm T_{eq}$ = 751 K, R = 2.12 $\rm R_\oplus$, \citealt{2021AJ....162..118E}), may have a cloudy atmosphere containing molecules heavier than hydrogen\citep{2014Natur.505...69K, 2014ApJ...794..155K}.
Water vapor discoveries have been reported in the atmosphere of two habitable zone super-Earths, LHS 1140 b ($\rm T_{eq}$ = 235 K, R = 1.7 $\rm R_\oplus$) and K2-18 b ($\rm M_p$ = 7.96 $\pm$ 1.91 $\rm M_\oplus$, $\rm R_p$ = 2.279 $\pm$ 0.0026 $\rm R_\oplus$) \citep{2021AJ....161...44E, 2019NatAs...3.1086T}. 
However, the exact nature of the atmospheres of some of these exoplanets mentioned above is still highly debated \citep[for example]{zhang21}.
With some upcoming astronomical observing facilities (e.g., JWST and Ariel), more constraints will be put on the atmospheric properties of terrestrial exoplanets. 

L~98-59 is a nearby (10.6 pc) bright (K = 7.1 mag) M3 dwarf star. Three terrestrial exoplanets have been discovered by the Transiting Exoplanet Survey Satellite (TESS) orbiting L~98-59, with a radius ranging from 0.8 to 1.6 $\rm R_\oplus$ \citep{2019AJ....158...32K, 2019A&A...629A.111C}. Their orbital periods are very short, ranging from 2.25 to 7.45 days.
As the innermost one in this multi-planet system, L~98-59~b has the lowest-mass by far measured for any exoplanet using the radial velocity technique \citep{2021A&A...653A..41D}.
\citet{2021AJ....162..169P} have evaluated the detectability of  molecular features in the planetary atmosphere with HST and JWST using simulated transmission spectra. 

Here we present an atmospheric study of L~98-59~b by analyzing the transmission spectral data obtained with the G141 grism of HST/WFC3. The outline of this study is as follows. We first describe the raw data reduction process using the Iraclis pipeline in Sec~\ref{sec:analysis} and how we retrieve the atmospheric properties of L~98-59~b using the TauRex code in Sec~\ref{sec:retrieve}. Then in Sec~\ref{sec:results}, we present our main analysis results and discussion. Finally, we give our conclusion in Sec~\ref{sec:conclusion}.

\section{Data Reduction Analysis} \label{sec:analysis}

The spatially scanned transiting observations of L~98-59~b taken between February 9th, 2020 and Feb 24th, 2021 using the HST/WFC3 G141 infrared grism (1.088 - 1.68 $\mu m$), are downloaded from the public Mikulski Archive for Space Telescopes (HST proposal 15856, PI: Thomas Barclay).

There were a total of five transits observed for L 98-59 b during that period. 
The observations were acquired using the 512 $\times$ 512 sub-array and the SPARS25 sample sequence with an exposure time of 69.62~s. With a scan rate of 0.496~$''$/$s$, the total scan length was 37.919~$''$.

We reduce these transit data using Iraclis, an open source pipeline designed and validated for the reduction of HST/WFC3 spatially scanning data (\citealp{2016ApJ...832..202T, 2016ApJ...820...99T} ) and available at GitHub\footnote{\url{https://github.com/ucl-exoplanets/Iraclis}}. 
The following steps were used for the reduction of the raw spectroscopic data with Iraclis: zero-read subtraction, reference pixel correction, non-linearity correction, dark current subtraction, gain conversion, sky background subtraction, calibration, flat-field correction, bad pixels, and cosmic-ray correction. Please see \citet{2016ApJ...832..202T, 2016ApJ...820...99T, 2018AJ....155..156T} for more details regarding these reduction steps. 

\begin{figure}
\gridline{\fig{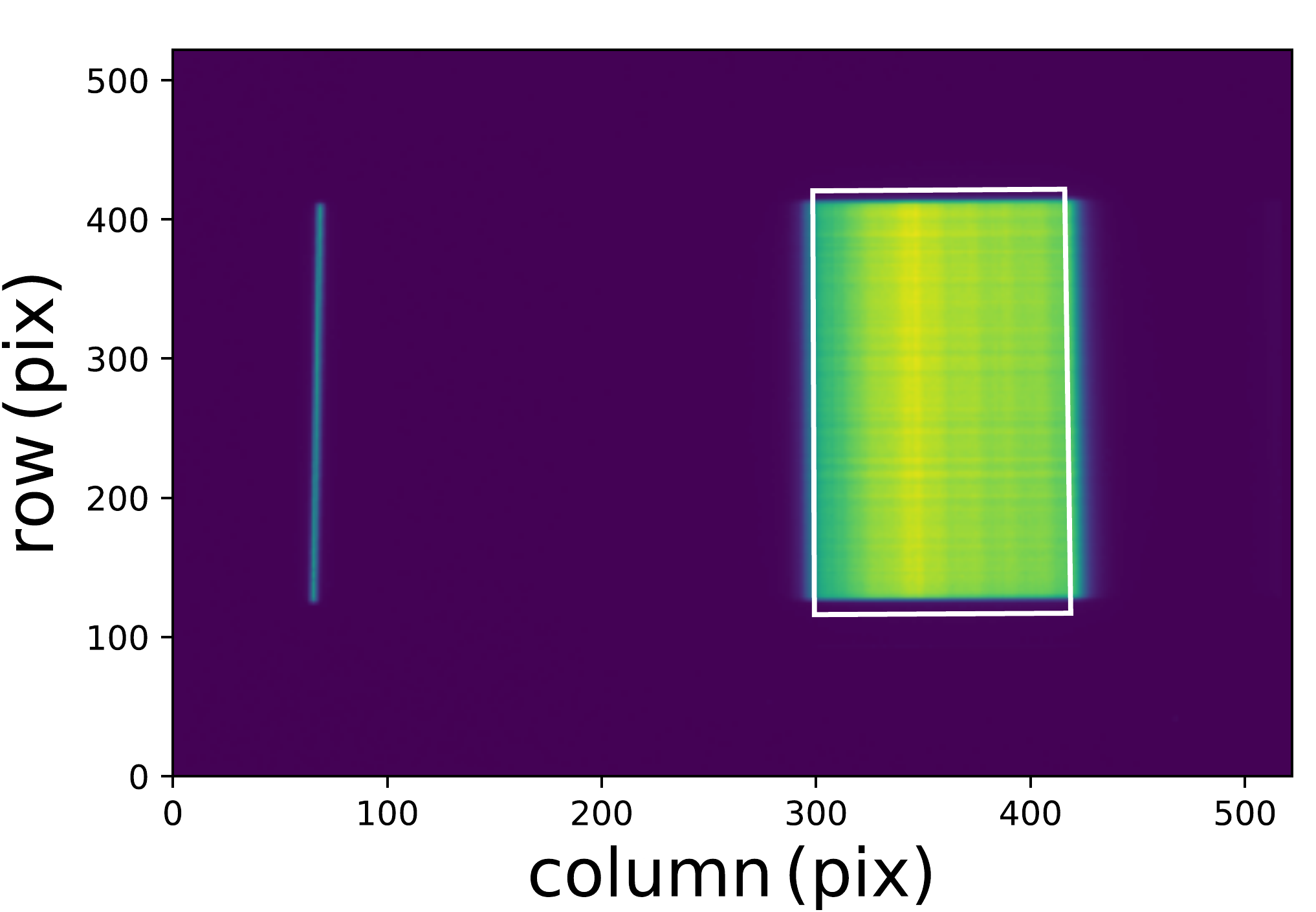}{0.5\textwidth}{(a)}}
\gridline{\fig{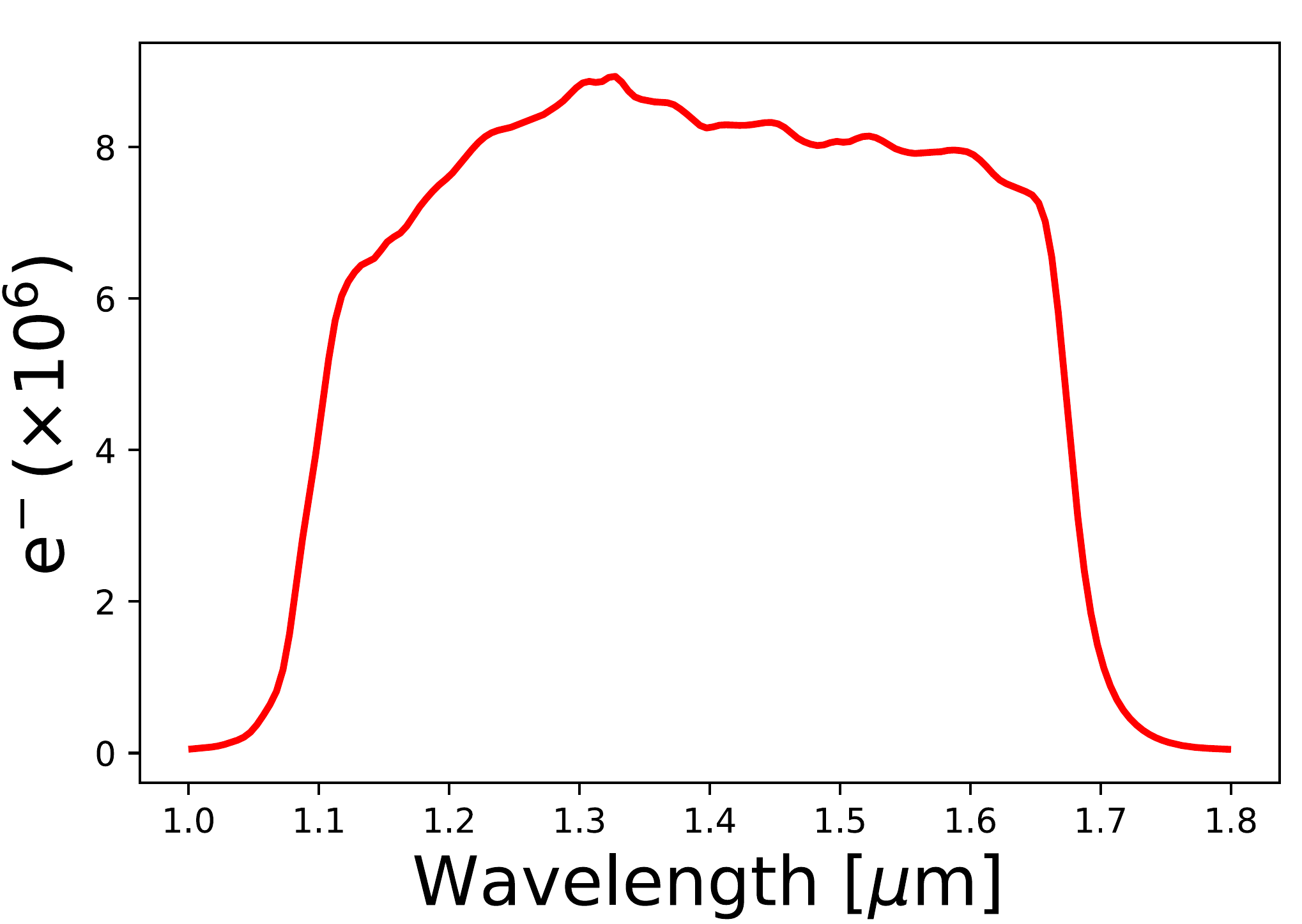}{0.5\textwidth}{(b)}}
\caption{(a) The detector image of forward scan in the first visit. The zeroth order and first order can be seen, from which we extract the spectrum . Flux in the white region is estimated. The region in the white square is photometric aperture. 
(b) Extracted 1D spectrum using Iraclis.}
\label{extraction}
\end{figure}

\begin{figure}
\gridline{\fig{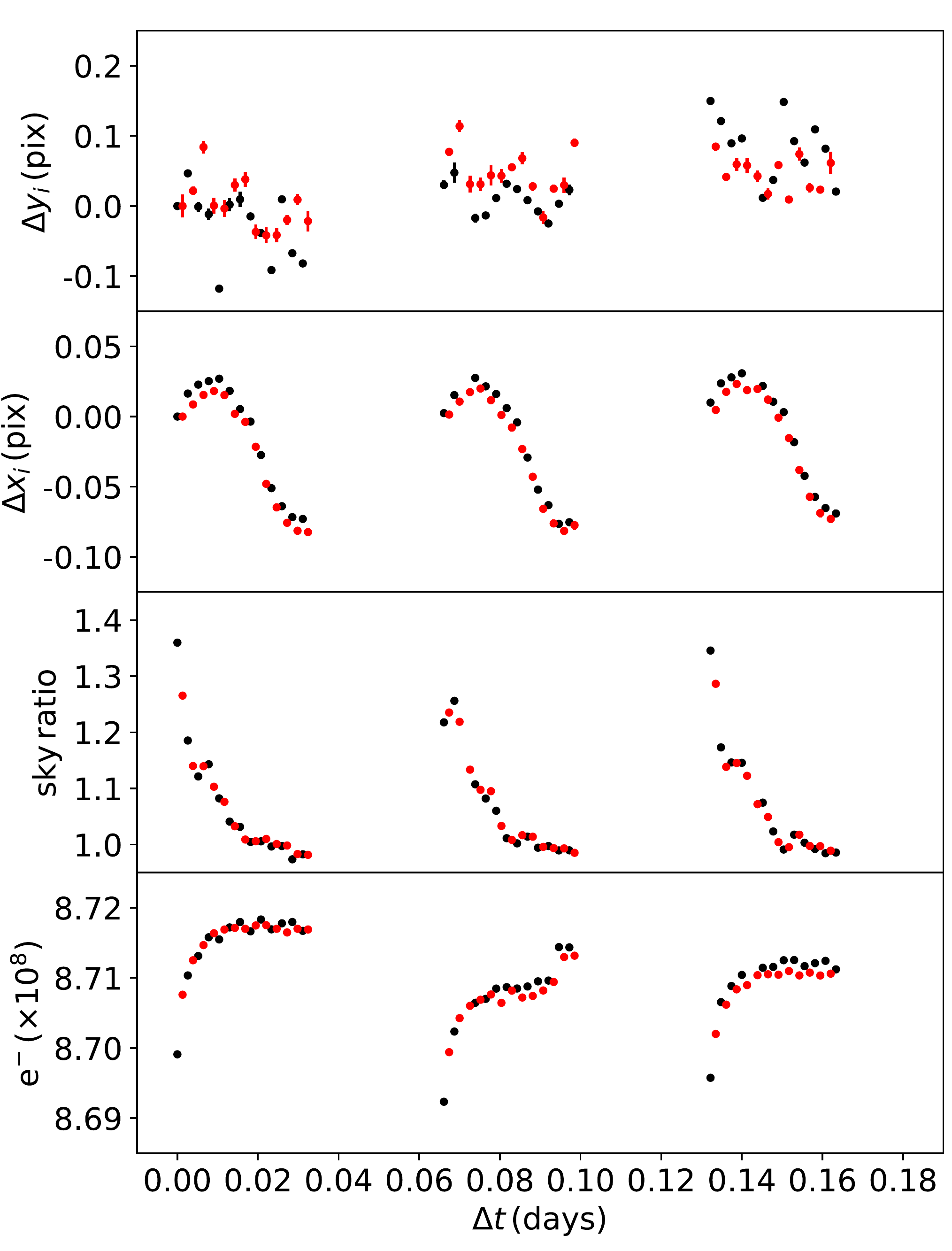}{0.5\textwidth}{}}
\caption{Vertical ($\Delta y$) and horizontal ($\Delta x$) shifts for each image relative to the first one, the sky ratio and raw flux of each image in the visit. Black: forward scan. Red: reverse scan. Here we take the first visit for example.}
\label{diagnostics}
\end{figure}

Then we extract the white and spectral light curves from the reduced spectroscopic data. The extraction region and extracted 1D spectrum are presented in Figure~\ref{extraction}. The specific geometric distortions of the spectroscopic images caused by the motion of the instrument in the scanning mode were taken into account. The position shifts are shown in Figure~\ref{diagnostics}. 
In our analysis, we exclude the first orbit of each transit visit because of a stronger wavelength-dependent ramp. The limb-darkening coefficients are calculated from a non-linear limb-darkening model \citep{2000A&A...363.1081C} using the stellar parameters of L~98-59 and the Phoenix stellar model \citep{2018A&A...618A..20C}. 
The stellar parameters used in the light curve fitting are taken from \citet{2021A&A...653A..41D} and are listed in Table~\ref{parameters}. We present the limb-darkening coefficients in Table~\ref{spectrum}. 

\begin{table}[t]
\centering
\caption{Parameters of L~98-59~b taken from \citet{2021A&A...653A..41D}, which are applied in Iraclis and TauRex.} \label{parameters}
\begin{threeparttable}
\resizebox{0.5\textwidth}{!}{
\begin{tabular}[b]{cc}
\hline\hline
\multicolumn{2}{c}{Stellar Parameters}\\
\hline
 $\rm [Fe/H](dex)$  & -0.46 $\pm$ 0.26\\
 $\rm T_{eff}(K)$ & 3415 $\pm$ 135\\
 $\rm M_*(M_{\odot})$ & 0.273 $\pm$ 0.030\\
 $\rm R_*(R_\odot)$ & 0.303 $^{+0.026}_{-0.023}$\\
 $\rm log(g_*)(cgs)$ & 4.86 $\pm$ 0.13\\
\hline\hline
\multicolumn{2}{c}{Planetary  Parameters}\\
\hline
$\rm T_{eq}(K)$ & 627 $^{+33}_{-36}$ \\
$\rm M_p(M_\oplus)$ & 0.40 $^{+0.16}_{-0.15}$\\
$\rm R_p(R_\oplus)$ & 0.85 $^{+0.061}_{-0.047}$\\
\hline\hline
\multicolumn{2}{c}{Transit  Parameters}\\
\hline
$\rm T_0(BJD)$ & 2458366.17067 $^{+0.00036}_{-0.00033}$ \\
$\rm P(days)$ & 2.2531136 $^{+0.0000012}_{-0.0000015}$\\
$\rm R_p/R_*$ & 0.02512 $^{+0.00072}_{-0.00064}$ \\
$\rm a(AU)$ & 0.02191 $^{+0.00080}_{-0.00084}$ \\
$\rm a/R_*$ & 15.0 $^{+1.4}_{-1.0}$ \\
$\rm i(deg)$ & 87.71 $^{+1.16}_{-0.44}$ \\
\hline
\hline
\end{tabular}}
\end{threeparttable}
\end{table}

\begin{table*}
\centering
\caption{The transit depths in different wavelength bins. \label{spectrum}}
\begin{threeparttable}
\resizebox{\textwidth}{!}{
\begin{tabular}[b]{cccccccc}
\hline
\multicolumn{8}{c}{\multirow{2}*{$(R_p/R_*)^2 (ppm)$}} \\
& & & & & & &\\
\hline
$\lambda_1$ ($\mu m$) & $\lambda_2$ ($\mu m$) & Visit 1 & Visit 2 & Visit 3 & Visit 4 & Visit 5 & Weighted Average \\
\hline
1.1153&1.1372&645.25 $\pm$ 58&595.33 $\pm$ 53&709.06 $\pm$ 56&598.0 $\pm$ 47&684.51 $\pm$ 51&646.25 $\pm$ 22\\
1.1372&1.1583&679.39 $\pm$ 45&624.83 $\pm$ 41&689.21 $\pm$ 37&548.58 $\pm$ 46&567.37 $\pm$ 45&633.52 $\pm$ 22\\
1.1583&1.1789&751.43 $\pm$ 44&689.85 $\pm$ 44&688.7 $\pm$ 40&642.95 $\pm$ 40&627.01 $\pm$ 47&681.32 $\pm$ 22\\
1.1789&1.1987&658.37 $\pm$ 50&680.27 $\pm$ 45&724.22 $\pm$ 38&613.01 $\pm$ 41&743.39 $\pm$ 44&683.09 $\pm$ 22\\
1.1987&1.2180&683.83 $\pm$ 43&646.41 $\pm$ 44&718.05 $\pm$ 44&694.1 $\pm$ 39&739.68 $\pm$ 48&695.87 $\pm$ 22\\
1.2180&1.2370&677.88 $\pm$ 41&608.73 $\pm$ 41&731.76 $\pm$ 40&731.5 $\pm$ 40&581.93 $\pm$ 44&669.70 $\pm$ 22\\
1.2370&1.2559&629.98 $\pm$ 42&650.36 $\pm$ 44&643.78 $\pm$ 35&662.38 $\pm$ 46&761.4 $\pm$ 37&672.26 $\pm$ 22\\
1.2559&1.2751&665.05 $\pm$ 37&588.46 $\pm$ 43&674.44 $\pm$ 40&720.14 $\pm$ 34&699.16 $\pm$ 54&671.20 $\pm$ 22\\
1.2751&1.2944&765.4 $\pm$ 41&627.03 $\pm$ 39&668.02 $\pm$ 39&732.07 $\pm$ 42&694.81 $\pm$ 43&691.18 $\pm$ 22\\
1.2944&1.3132&651.22 $\pm$ 38&610.17 $\pm$ 38&689.57 $\pm$ 38&660.51 $\pm$ 41&762.37 $\pm$ 46&667.14 $\pm$ 22\\
1.3132&1.3320&688.28 $\pm$ 38&598.81 $\pm$ 45&737.05 $\pm$ 41&685.85 $\pm$ 45&724.01 $\pm$ 41&684.70 $\pm$ 22\\
1.3320&1.3509&672.42 $\pm$ 40&615.58 $\pm$ 37&705.26 $\pm$ 43&613.21 $\pm$ 41&712.07 $\pm$ 41&665.02 $\pm$ 22\\
1.3509&1.3701&662.96 $\pm$ 49&578.78 $\pm$ 43&703.64 $\pm$ 41&713.9 $\pm$ 46&707.76 $\pm$ 47&675.52 $\pm$ 22\\
1.3701&1.3900&686.03 $\pm$ 40&566.08 $\pm$ 38&748.49 $\pm$ 35&713.52 $\pm$ 41&705.19 $\pm$ 53&680.49 $\pm$ 22\\
1.3900&1.4100&714.95 $\pm$ 38&625.51 $\pm$ 44&738.81 $\pm$ 38&648.56 $\pm$ 37&741.06 $\pm$ 40&697.93 $\pm$ 22\\
1.4100&1.4303&677.26 $\pm$ 44&650.65 $\pm$ 44&775.89 $\pm$ 33&676.22 $\pm$ 39&735.16 $\pm$ 40&712.00 $\pm$ 22\\
1.4303&1.4509&624.61 $\pm$ 43&679.29 $\pm$ 38&686.32 $\pm$ 33&652.23 $\pm$ 30&681.06 $\pm$ 41&668.31 $\pm$ 22\\
1.4509&1.4721&718.08 $\pm$ 40&657.39 $\pm$ 44&674.02 $\pm$ 39&650.47 $\pm$ 39&728.44 $\pm$ 51&685.98 $\pm$ 22\\
1.4721&1.4941&702.75 $\pm$ 39&652.01 $\pm$ 39&740.52 $\pm$ 34&659.62 $\pm$ 35&672.29 $\pm$ 41&689.61 $\pm$ 22\\
1.4941&1.5165&640.03 $\pm$ 42&642.43 $\pm$ 46&754.34 $\pm$ 34&596.47 $\pm$ 40&705.65 $\pm$ 43&675.71 $\pm$ 22\\
1.5165&1.5395&703.8 $\pm$ 40&734.88 $\pm$ 48&740.23 $\pm$ 39&602.52 $\pm$ 33&680.89 $\pm$ 42&681.45 $\pm$ 22\\
1.5395&1.5636&704.98 $\pm$ 37&797.13 $\pm$ 44&710.15 $\pm$ 36&704.32 $\pm$ 35&710.06 $\pm$ 49&718.87 $\pm$ 22\\
1.5636&1.5889&735.58 $\pm$ 37&693.72 $\pm$ 41&747.85 $\pm$ 39&744.42 $\pm$ 43&680.77 $\pm$ 40&720.09 $\pm$ 22\\
1.5889&1.6153&628.66 $\pm$ 38&630.33 $\pm$ 34&710.26 $\pm$ 34&621.54 $\pm$ 34&624.5 $\pm$ 42&646.05 $\pm$ 22\\
1.6153&1.6436&632.81 $\pm$ 35&623.13 $\pm$ 36&675.7 $\pm$ 38&642.63 $\pm$ 34&686.16 $\pm$ 48&650.16 $\pm$ 22\\
\hline
\end{tabular}}
\end{threeparttable}
\end{table*}

\begin{figure*}
\gridline{\fig{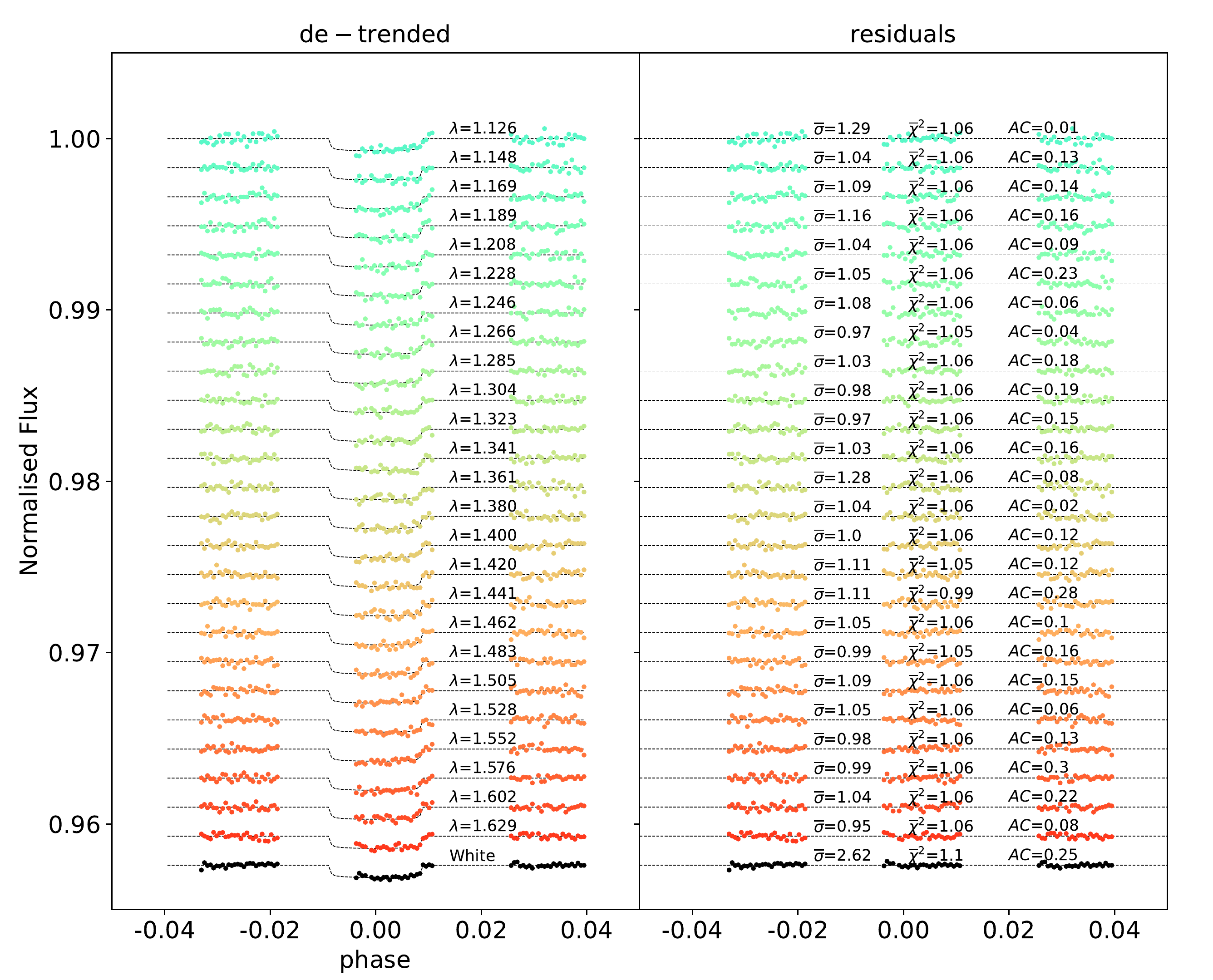}{1.0\textwidth}{}}
\caption{Spectral light curves for the transmission spectra. Left panel: the detrended spectral light curves with best-fit model
plotted. The right panel reveals the residuals. For each spectra, we present the values of $\sigma$, chi-square ($\chi^2$) and the
AC.}
\label{all_fitting}
\end{figure*}

\begin{figure*}
\includegraphics[width=1.0\textwidth]{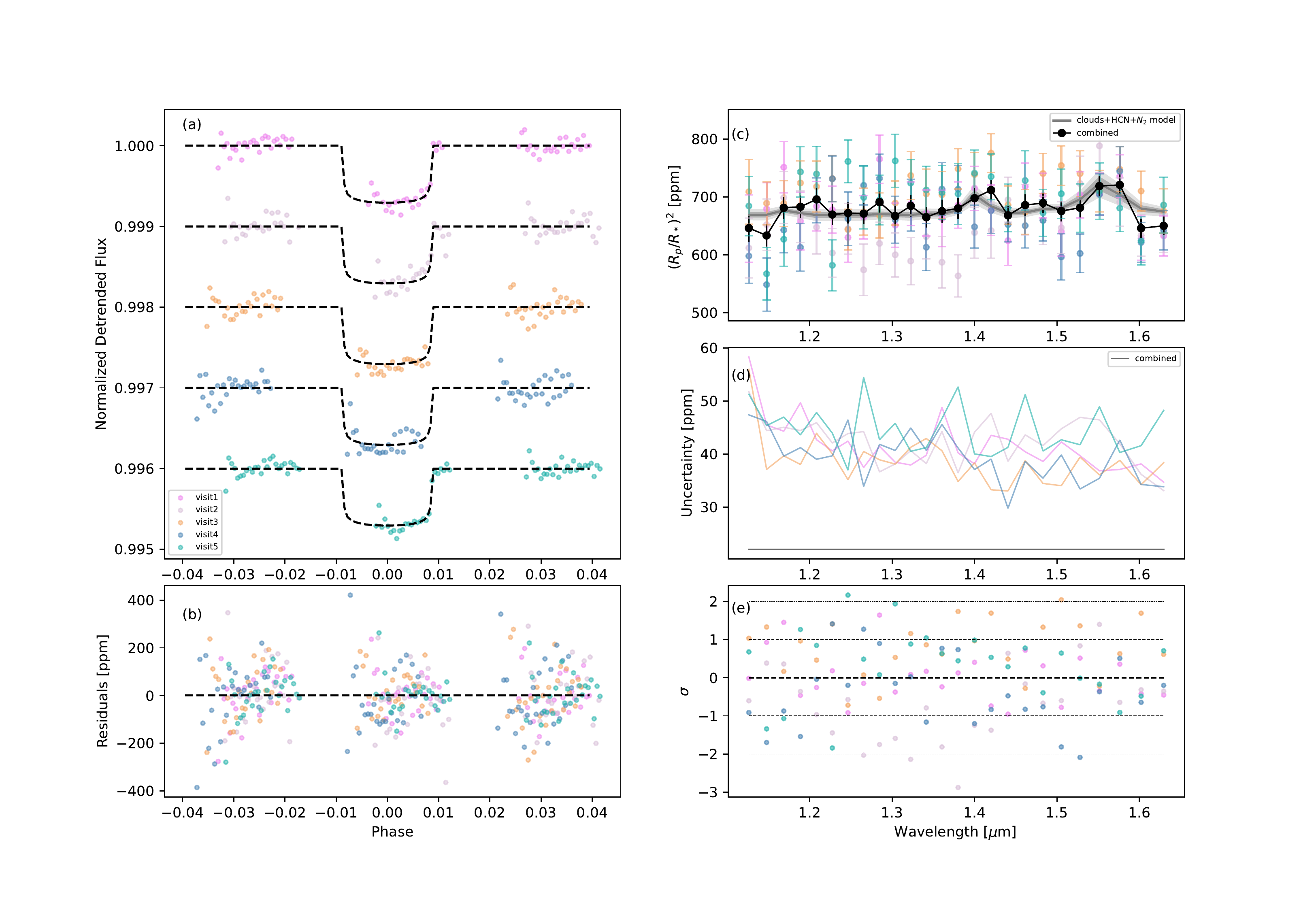}
\caption{The panels (a) and (b) show the white light curves and residuals for best-fit models of 5 individual transit observations, respectively. Panel (c): transmission spectrum collected from each visit (colored data points) and the final combined transmission spectrum (black data points). Panel (d): uncertainty of each data point. Panel (e): deviation from the weighted-average spectrum, in unit of sigma, for each data point. }
\label{lcs_iraclis}
\end{figure*}

When fitting the white light curves with a transit model, we set the mid-transit time and the ratio between the planet radius and the host star radius as free parameters, while other stellar, planetary, and orbital parameters are fixed. In the second observation, we exclude two defect points with large deviations from the whole. In panels a and b of Figure~\ref{lcs_iraclis}, we present the white light curves together with the best fitting model and residuals for the 5 transit visits. 
Then we fitted spectral light curves with a fixed mid-transit time derived from white light curve fitting and set the planet and stellar radius ratio as free parameters. The spectral light curve fits are shown in Figure~\ref{all_fitting}, where an offset has been applied for clarity.

The spectra of 5 individual visits and the final combined spectrum are shown in the right panel of Figure~\ref{lcs_iraclis}. 
The final combined transmission spectrum was obtained by calculating a weighted average of five transits spectra. Before combining the five visits into a weighted-average spectrum, we checked that the fitted transit depths in each bin are drawn from the same distribution and there are no significant outliers. 
Compared with the combined spectrum, most of the data points from individual visits demonstrate $<2\sigma$ spread. And removing the outlying visit data in these spectrum bins has no appreciable effect on the final spectrum. We therefore keep all five visits when calculating the average transit depth for each wavelength bin. 
The final measurement uncertainty across the WFC3/G141 bandpass was estimated by calculating the standard deviation of the residuals when subtracting the combined spectrum from spectrum of individual visit, which is $\sim$22~ppm. Therefore, we will take 22~ppm as the precision of the final combined transmission spectrum in our remaining analysis.

\section{Atmospheric Modeling} \label{sec:retrieve}

To retrieve the atmospheric parameters of L~98-59~b, we use the most recent version of the public available code  TauRex3\footnote{\url{https://github.com/ucl-exoplanets/TauREx3_public}} \citep{2021ApJ...917...37A, 2015ApJ...813...13W, 2015ApJ...802..107W} . Based on a Bayesian atmospheric retrieval framework, TauRex3 maps the parameter space of a planetary atmospheric model and finds the best fit to the observed spectrum using the nested sampling code MultiNest \citep{2014A&A...564A.125B}. 
For our analysis, we set the live points number to 1500 and the tolerance evidence to 0.5 in the retrieval code.
Due to the narrow wavelength coverage of our observed data, we typically only probe a very restricted range of the planet’s temperature–pressure profile.
Here we adopted a planetary atmosphere model with an isothermal temperature-pressure profile, a constant molecular abundance at each layer of the atmosphere, and 100 plane-parallel layers with pressure ranging from $10^{-3}$ to $10^7$~Pa. We select two uniform priors, 0.6~$\rm T_{eq}$ to 1.6~$\rm T_{eq}$ for planet temperature and 0.5~$R_p$ to 1.5~$R_p$ for planet radius, where $\rm T_{eq}=627~K$ is the planet's equilibrium temperature.

For the opacity sources, we initially assume the atmosphere is dominated by H$_2$ and $\rm He$ with the ratio fixed to $\rm He/H_2=0.17$. 
Then, to model the transmission spectrum, we introduce additional active trace-gases, including $\rm HCN$ \citep{2014MNRAS.437.1828B}, $\rm CH_4$ \citep{2014MNRAS.440.1649Y}, $\rm NH_3$ \citep{2019MNRAS.490.4638C}, $\rm CO$ \citep{2015ApJS..216...15L}, $\rm CO_2$ \citep{2010JQSRT.111.2139R}, $\rm H_2O$ \citep{2018MNRAS.480.2597P}, $\rm TiO$ \citep{2019MNRAS.488.2836M}, and $\rm VO$ \citep{2016MNRAS.463..771M}. The line lists are taken from ExoMol \citep{2016JMoSp.327...73T,2021A&A...646A..21C}, HITEMP \citep{2010JQSRT.111.2139R} and HITRAN \citep{1987ApOpt..26.4058R}. 
All of our models include collision-induced absorption (CIA) of $\rm H_2$-$\rm H_2$ \citep{doi:10.1021/jp109441f,2018ApJS..235...24F} and $\rm H_2$-$\rm He$ \citep{doi:10.1063/1.3676405}, Rayleigh scattering. 
Clouds are modeled as gray clouds \citep{2013ApJ...778...97L}, with the cloud top pressure ranging from $10^{-2}$ to $10^7$~Pa. 
$\rm  N_2$ is an inactive gas in the atmosphere over the HST/WFC3 spectral wavelength range, which only contributes to the mean molecular weight. 
Thus, to evaluate whether L~98-59~b can have a heavy secondary atmosphere \citep{2021AJ....161..284M}, we will add $\rm  N_2$ \citep{2018JQSRT.219..127W} to some of the atmospheric models later. 

We run two initial retrievals aiming at identifying the most likely trace gases, one in which the trace-gas abundance is set to vary from $10^{-12}$ to $10^{-1}$ in volume mixing ratios (VMR), and the other one from $10^{-12}$ to 1. This will allow a larger degree of freedom when exploring the parameter space of active trace gases abundances. 
Based on the results of two initial fittings, HCN is the only actively absorbing molecule that emerges. Therefore, we choose to evaluate five atmospheric models in our second round of retrieval in Sec~\ref{sec:results}, four with HCN and one without HCN: a cloud-free atmosphere containing only HCN (model 1), a cloud-free atmosphere containing HCN and $\rm N_2$ (model 2), an atmosphere containing clouds and HCN (model 3), an atmosphere containing clouds, HCN and $\rm N_2$ (model 4), and a pure cloudy atmosphere (model 5).

\section{Results and Discussion} \label{sec:results}

\begin{table*}
\centering
\caption{Statistical values of different models: Bayesian evidences, $\chi^2$, ADI and Sigma. }
\label{retrieval_results}
\begin{threeparttable}
\resizebox{\textwidth}{!}{
\begin{tabular}[b]{lllllll}
\hline
&&&&&\\
&&HST&&&\\
&&&&&\\
\hline
model & descriptions & VMR priors & $\rm log_{10}$(E) & $\chi^2$ & ADI & Sigma\\
\hline
model 1 & HCN & $10^{-12}$ $\sim$ 1 & 227.025 $\pm$ 0.071 & 18.583 & - & -\\
\hline
model 2 & HCN+$\rm N_2$ & $10^{-12}$ $\sim$ 1 & 227.925 $\pm$ 0.068 & 16.628 & - & -\\
\hline
model 3 & clouds+HCN & $10^{-12}$ $\sim$ 1 & 229.563 $\pm$ 0.065 & 14.028 & 1.023 & 2.04$\sigma$\\
\hline
model 4 & clouds+HCN+$\rm N_2$ & $10^{-12}$ $\sim$ 1 & 229.740 $\pm$ 0.064 & 14.025 & 1.2 & 2.14$\sigma$\\
\hline
model 5 & clouds & - & 228.540 $\pm$ 0.055 & 24.131 & - & -\\
\hline
\hline
&&&&&\\
&&HST+TESS&&&\\
&&&&&\\
\hline
model 4 & clouds+HCN+$\rm N_2$ & $10^{-12}$ $\sim$ 1 &  238.019 $\pm$ 0.064 & 14.345 & 1.031 & 2.04 $\sigma$\\
\hline
model 5 & clouds & - & 236.988 $\pm$ 0.055 & 24.318 & - & -\\
\hline
\hline
\end{tabular}}
\end{threeparttable}
\end{table*}

\begin{table*}
\centering
\caption{The best fit model parameters of different models.}
\label{retrieval_parameters}
\begin{threeparttable}
\resizebox{\textwidth}{!}{
\begin{tabular}[b]{llllllll}
\hline
model & descriptions & VMR priors & $\rm R_p (R_{Jup})$ & $\rm T_{equ} (K)$ & log(P)(Pa) & $\rm \mu$ (amu)\\
\hline
model 4 & clouds+HCN+$\rm N_2$ & $10^{-12}$ $\sim$ 1 & 0.06$^{+0.01}_{-0.01}$ & 593.70$^{+199.8}_{-153.92}$ & 2.02$^{+1.76}_{-1.08}$ & 11.68$^{+9.21}_{-6.04}$\\
\hline
model 5 & clouds & - & 0.06$^{+0.01}_{-0.01}$ & 441.55$^{+97.94}_{-62.51}$ & 4.87$^{+1.45}_{-1.31}$ & 2.30$^{+0.00}_{-0.00}$\\
\hline
\hline
\end{tabular}}
\end{threeparttable}
\end{table*}

To evaluate the goodness of fitting and detection significance of an atmospheric model, we use $\chi^2$ value, Bayesian evidence, and the atmospheric detectability index \citep[ADI][]{2018AJ....155..156T}, which are summarized in Table~\ref{retrieval_results}.
The ADI is a positively defined Bayes factor \citep{doi:10.1080/01621459.1995.10476572} between models with active trace molecules and a pure cloudy model, where a value of greater than 3 means a significant detection. 
Table~\ref{retrieval_results} also includes the n-$\sigma$ values, where 3.0$\sigma$ indicates a moderate to strong detection \citep{2013ApJ...778..153B}. The n-$\sigma$ values are calculated with reference to Equation(10) and Equation(11) in \cite{2013ApJ...778..153B}.
The best-fit models are presented in Figure~\ref{models}, and parameter results are summarized in Table~\ref{retrieval_results} and Table~\ref{retrieval_parameters}. 

\subsection{Atmospheric Models with HCN}
\label{sec:model_with_HCN}
\begin{figure}
\includegraphics[width=0.5\textwidth]{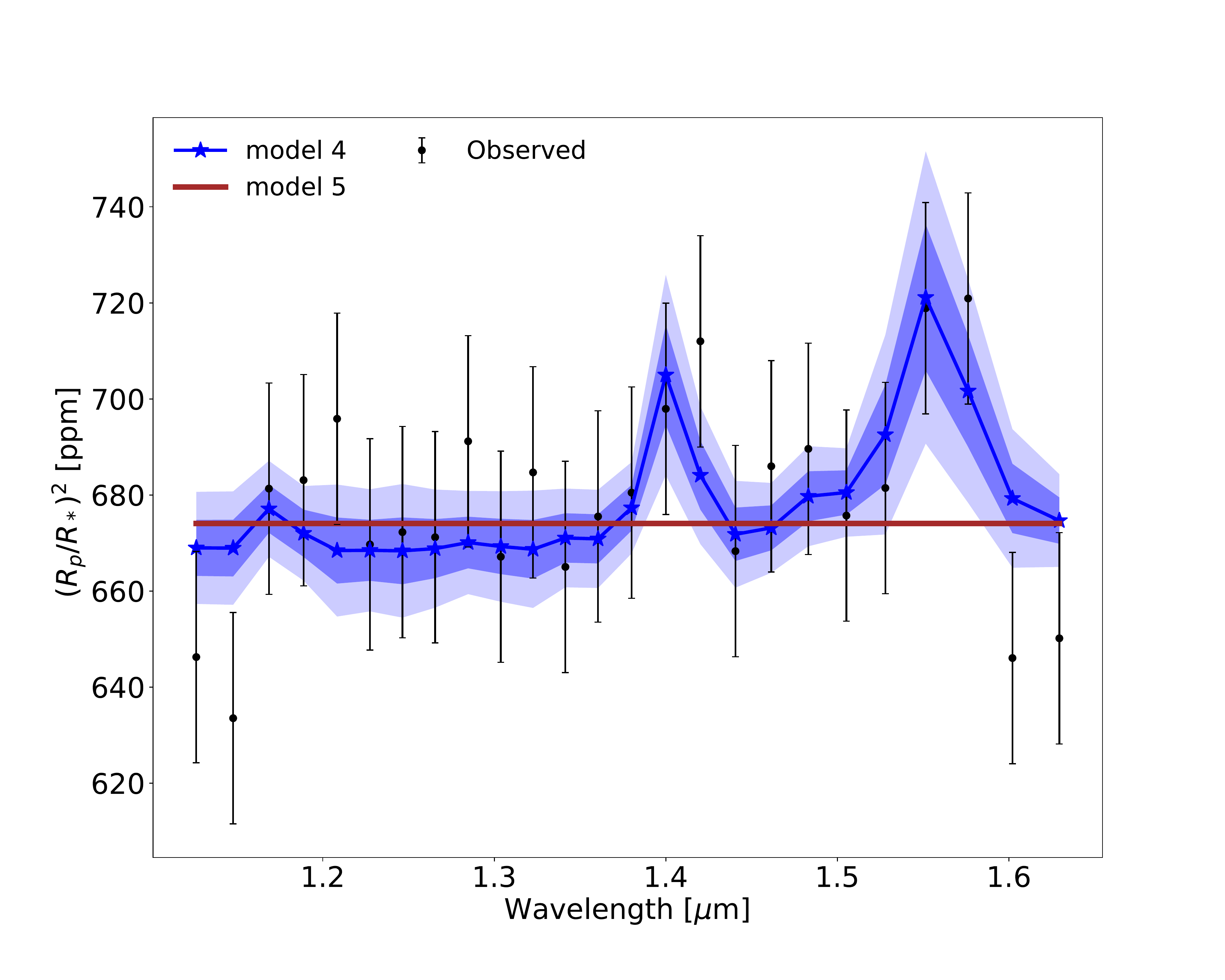}
\caption{The spectrum of different atmospheric models of HST/WFC3.}
\label{models}
\end{figure}

By visually inspecting the combined spectrum, we find two weak absorption features centered near 1.40 and 1.55 $\mu m$ in the transmission spectrum of L~98-59~b shown in Figure~\ref{models}. These features can be explained by the presence of HCN in its atmosphere, which is also supported by the initial retrieval results.
For all the models with HCN included (models 1, 2, 3 and 4), the model with clouds, HCN and $\rm N_2$ (model 4) yields the highest Bayesian evidence and ADI value, which is also shown Figure~\ref{lcs_iraclis} and \ref{models}, with 1$\sigma$ (shadow regions with heavier gray) and 2$\sigma$ (shadow regions with lighter gray) uncertainties over-plotted.

\begin{figure*}
\gridline{\fig{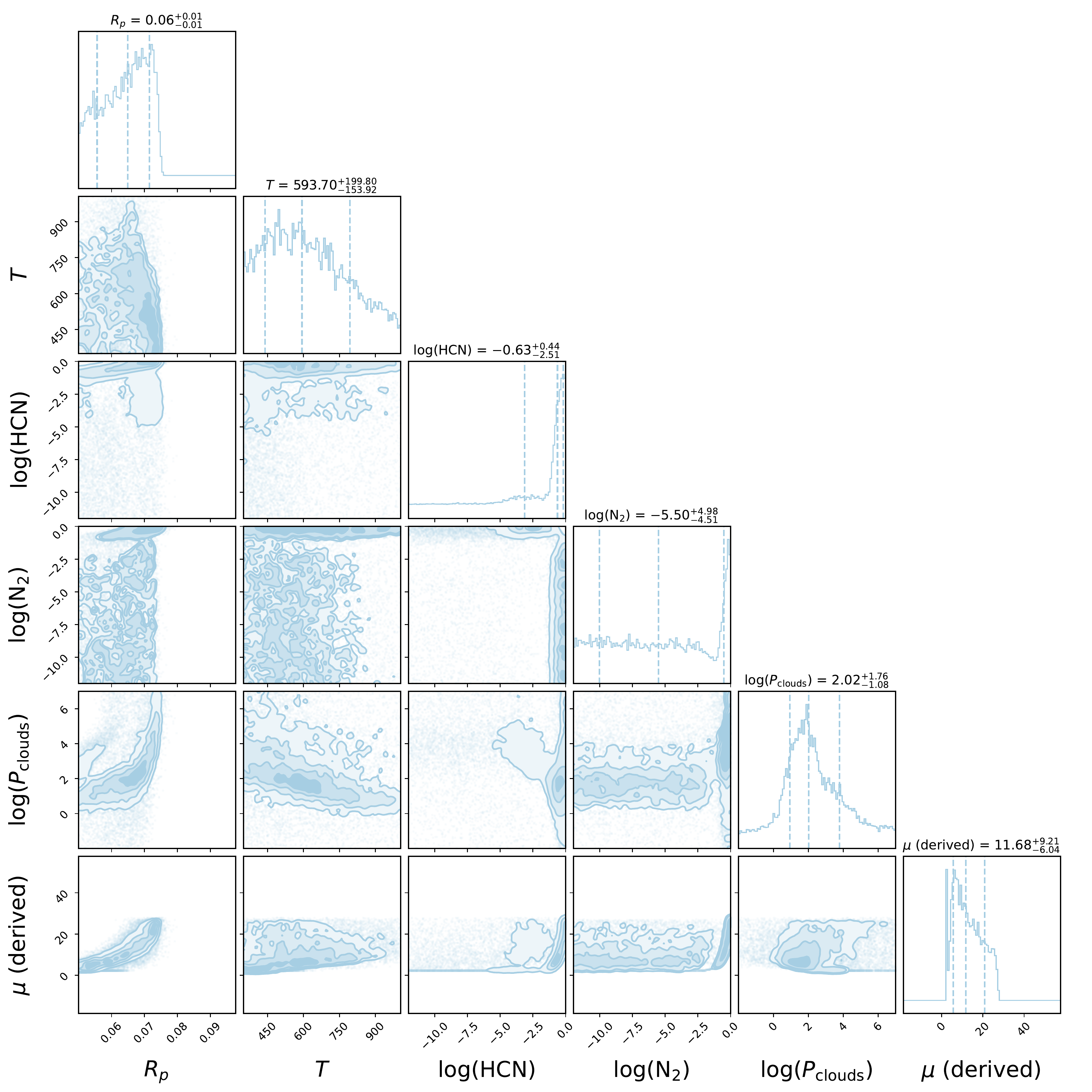}{1.0\textwidth}{}}
\caption{The atmospheric retrieval parameters of the best-fit model (model 4) for L~98-59~b.}
\label{taurex-posterios-best}
\end{figure*}

Figure~\ref{taurex-posterios-best} presents the posterior distributions of the retrieved parameters from this model 4, such as planet radius, equilibrium temperature, molecular abundance mixing ratios, surface pressure of the gray cloud, and mean molecular weight. 
The scale height of the atmosphere is about 49~km ($\sim12$~ppm) with $\rm T = 593.7$~K, $\mu$ = 11.68~amu, and $R_p$ = 0.06~$R_{\rm Jup}$. 
The posterior distribution of the surface pressure, which is the pressure at the top of a fully opaque cloud deck, peaks around $10^{-3}$~bar. 
The abundance of $\rm HCN$ is $10^{-0.63^{+0.44}_{-2.51}}$, which is much higher than the HCN abundance results from super-Earth 55 Cancri e \citep{2016ApJ...820...99T} and GJ~1132~b \citep{2021AJ....161..213S}. 

To extend the wavelength coverage, we also ran atmosphere retrieval by including data from the TESS mission \citep{2021A&A...653A..41D}. The fitting results are shown in Table~\ref{retrieval_results}.
Among all the models with HCN included, model 4 still shows the highest statistical significance. However, according to \cite{2013ApJ...778..153B}, the difference in Bayesian evidence between different models is too small to give a conclusive result.

In Sec~\ref{equilibrium chemistry}, we will calculate the abundance of HCN using an equilibrium chemistry model. We will show that the high value of HCN abundance derived above cannot be explained by an equilibrium chemical model assuming a reasonable C/O ratio and metallicity, thus is likely to be nonphysical and should be discarded.

\begin{figure*}
\gridline{\fig{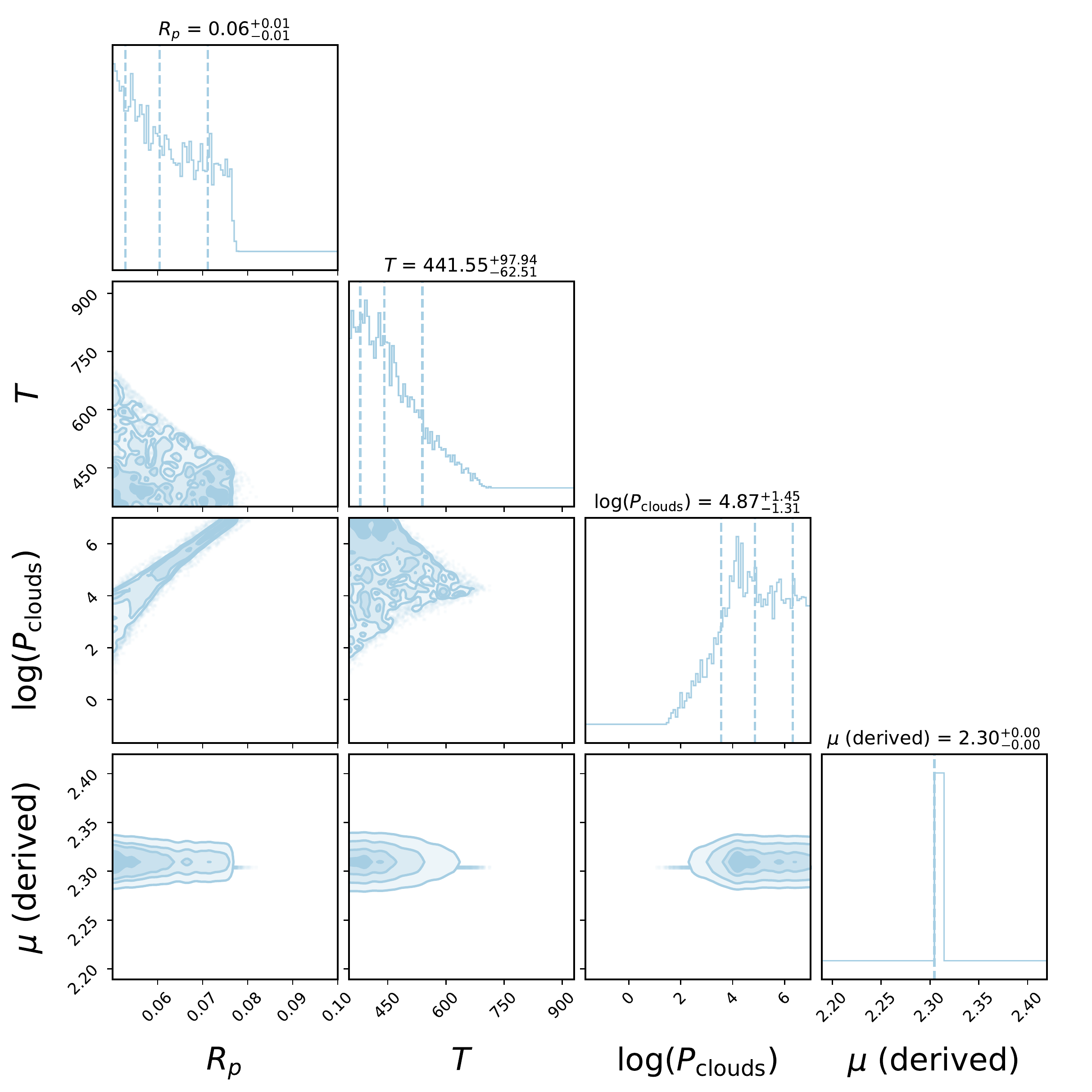}{1.0\textwidth}{}}
\caption{Same as Figure~\ref{taurex-posterios-best}, but for the flat-line pure cloudy model (model 5).}
\label{taurex-posterios-cloudy}
\end{figure*}

\subsection{HCN Abundance from Equilibrium Chemistry}
\label{equilibrium chemistry}

We find the transmission spectrum of L~98-59~b is best fitted by the model with abundant HCN in a high mean molecular weight atmosphere. 
HCN is a fundamental molecule in the origins of life. 
In atmospheres, HCN generally forms out of the reactive radicals left over from methane and nitrogen dissociation by ultraviolet light, galactic cosmic rays, and lightning \citep{catling17, horst17, pearce22}. 
The most HCN-rich atmosphere in the solar system belongs to Titan \citep[several parts per thousand in the upper atmosphere,][]{Koskinen11}.
Previously HCN has been detected in the atmosphere of the highly irradiated super-Earth 55 Cancri e \citep{2016ApJ...820...99T} and GJ~1132~b \citep{2021AJ....161..213S}.
Please also see \citet{2021AJ....161..284M} for their study on GJ~1132~b, in which no evidence was found for HCN absorption in the HST transmission spectrum. 
\citet{2021AJ....161..213S} explored several scenarios in which GJ~1132~b reestablished an atmosphere with HCN and CH4 after the loss of its primary atmosphere, which included mantle outgassing, volcanic activity caused by tidal heating, and lightening.
Because L~98-59~b is located in a packed planetary system and has an orbital eccentricity of 0.1 \citep{2021A&A...653A..41D}, gravitational forces from the central star and other planets can squeeze the planet and cause strong internal tidal heating. 
Thus, for L~98-59~b, it likely has lost its primary $\rm H_2$ dominated atmosphere due to photo-evaporation because of the close proximity to its central star. 
Then the high–molecular weight secondary atmosphere with HCN was revived by the volcanic and mantle outgassing processes from tidal heating. 

However, the HCN abundance derived in Sec~\ref{sec:model_with_HCN} seems really high. 
To check if such a high HCN abundance in the atmosphere of L~98-59~b is plausible, we apply the state-of-the-art chemical equilibrium code of FastChem \citep{2018MNRAS.479..865S, 2022arXiv220608247S} to calculate the gas phase chemical composition under thermochemical equilibrium approximation for a given temperature and pressure \footnote{\url{https://github.com/exoclime/FastChem}}.

\begin{figure}[htbp]
\gridline{\fig{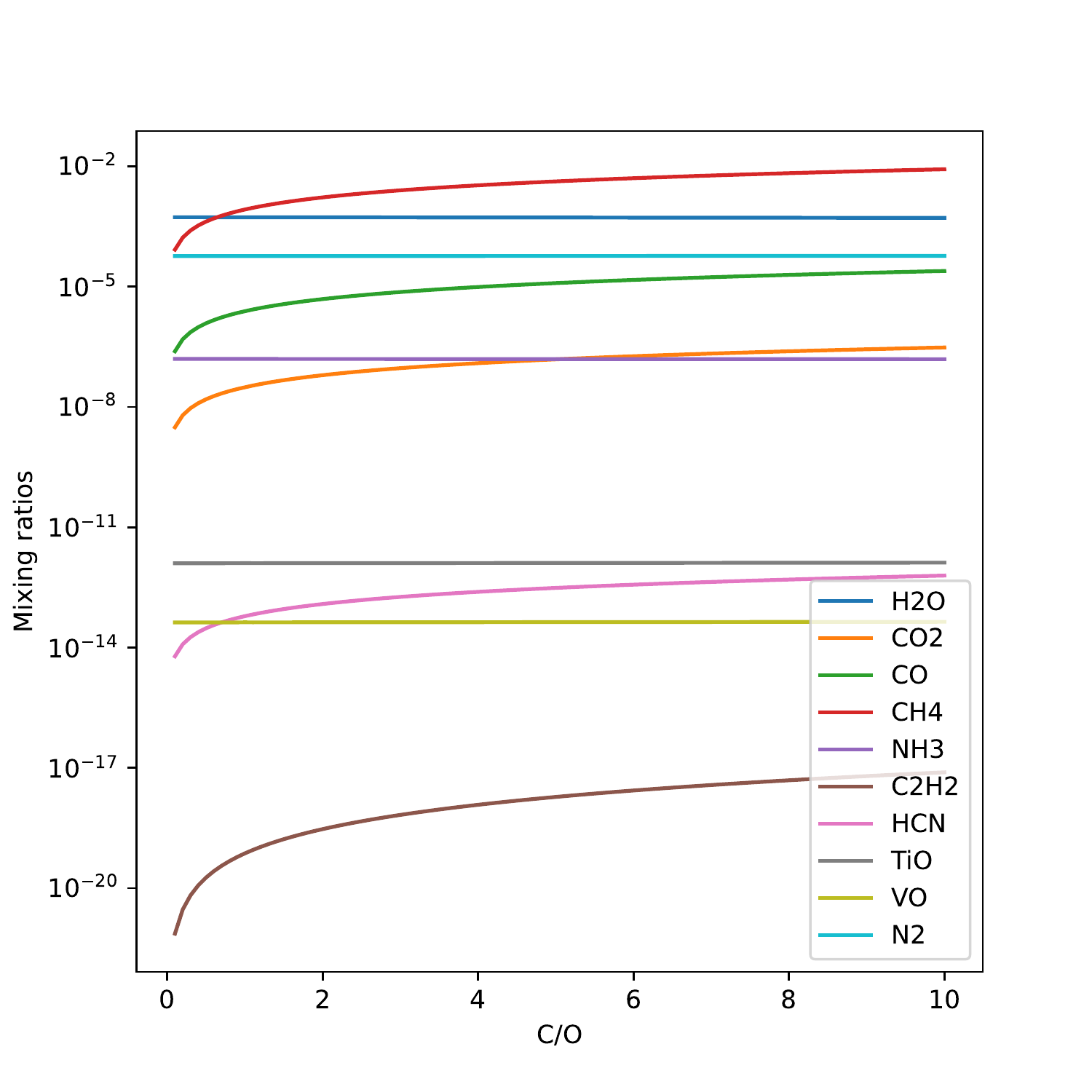}{0.5\textwidth}{}}
\caption{The abundance of molecules range with different C/O ratios using equilibrium chemistry model.}
\label{fastchem}
\end{figure}

\begin{figure}[htbp]
\gridline{\fig{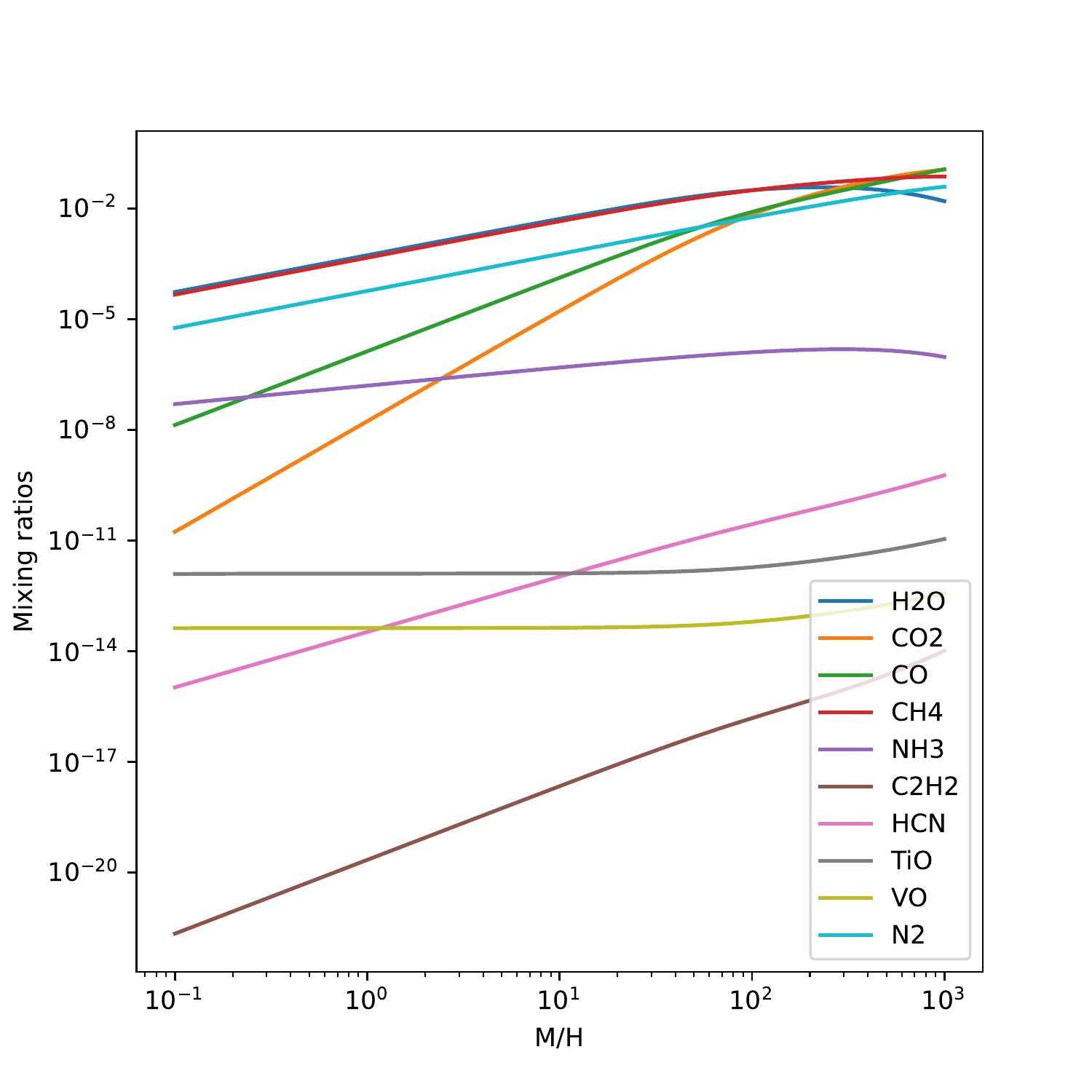}{0.5\textwidth}{}}
\caption{The abundance of molecules range with different metallicity (0.1 - 1000$\times$solar metallicity) using equilibrium chemistry model.}
\label{fastchem_metallicity}
\end{figure}

We set the temperature to 627~K and the pressure to 1~mbar. We first investigate the impact of C/O ratio on the HCN abundance by assuming solar photospheric abundances for all elements except for C and O  \citep{2009ARA&A..47..481A}. 
We present the calculated molecular abundance distributions in Figure~\ref{fastchem}. 
From this figure, we can see that the abundance of HCN varies between $10^{-14}$ and $10^{-12}$ when we change the C/O ratio from 0.1 to 10. 
We have also explored the impact of metallicity on the abundance of HCN, and the results are shown in Figure~\ref{fastchem_metallicity}.
We can see the VMR of HCN changes from $10^{-15}$ to $10^{-9}$ when the metallicity is increased from 0.1 to 1000 solar metallicity. 
Although the abundance of HCN will increase when there is more carbon available, it is still much smaller than those of other trace gases.  
Therefore we cannot explain the high abundance of HCN derived in Sec~\ref{sec:model_with_HCN} for L~98-59~b using an equilibrium chemistry model with a reasonable metallicity and C/O ratio.  
Therefore, we will seek an alternative model in Sec~\ref{model with cloud} to explain the transmission spectrum of L~98-59~b.

\subsection{Models With Clouds and Hazes}
\label{model with cloud}

Here we will explore atmospheric models without the active absorber of HCN. The best retrieval result under this scenario is a flat-line (or the so called ``pure cloudy") model.
Generally, flat and featureless exoplanet transmission spectra can be attributed to three different causes: a higher mean molecular weight atmosphere (and thus smaller scale height), a high-altitude aerosol layer blocking absorption features, or no atmosphere at all. 
While L~98-59~b’s temperature and terrestrial nature suggest a thin atmosphere with higher mean molecular weight, the WFC3 observations alone cannot rule out the possibility of
a H/He dominated atmosphere combined with condensate clouds or organic photochemical hazes (please see also our discussion in Sec~\ref{sec:loss}).
The best fitted model parameters are shown in Table~\ref{retrieval_results}, with the model curve also shown in Figure~\ref{lcs_iraclis} and \ref{models}. 

In Figure~\ref{taurex-posterios-cloudy} we present the posterior distributions of the retrieved parameters, such as planet radius, equilibrium temperature, surface pressure of the gray cloud, and mean molecular weight. The posterior distribution of the surface pressure, which is the pressure at the top of a fully opaque cloud deck, peaks around 1~bar. The mean molecular weight 2.3 is calculated using the fixed H$_2$ to He ratio.

For the pure cloudy model, the Bayesian evidence of the best fit is $\rm log(E)=228.54(0.055)$, a little lower than that of the HCN-only cloudy model, which has $\rm log(E)=229.740(0.064)$. According to \cite{2013ApJ...778..153B}, there is no statistical differences between these two models. Take the equilibrium chemical model results from Sec~\ref{equilibrium chemistry} into account, we ultimately determine that all evidence points to L~98-59~b having a featureless WFC3 spectrum, which can be explained by a thin atmosphere with high mean molecular weight, an atmosphere with highly opaque aerosol layer, or having no atmosphere. And these scenarios cannot be distinguished with the current observation data from WFC3.

It is likely that there is small non-Gaussian noise left unaccounted for in the spectral data extraction, which can cause inflation of the errors of final combined spectrum. Thus we have re-run the retrieval using inflated error bars (such as 25~ppm and 30~ppm) instead of the old 22~ppm to evaluate the robustness of the atmospheric retrieval. When using an error bar larger than 22~ppm, the Bayesian evidence of the flat-line pure cloudy model (model 5) becomes larger than the cloudy model with HCN and N$_2$ (model 4). This also supports our preference of using a flat-line pure cloudy model to explain the transmission spectrum of L~98-59~b.  

\subsection{Future Missions}
\begin{figure}
\includegraphics[width=0.5\textwidth]{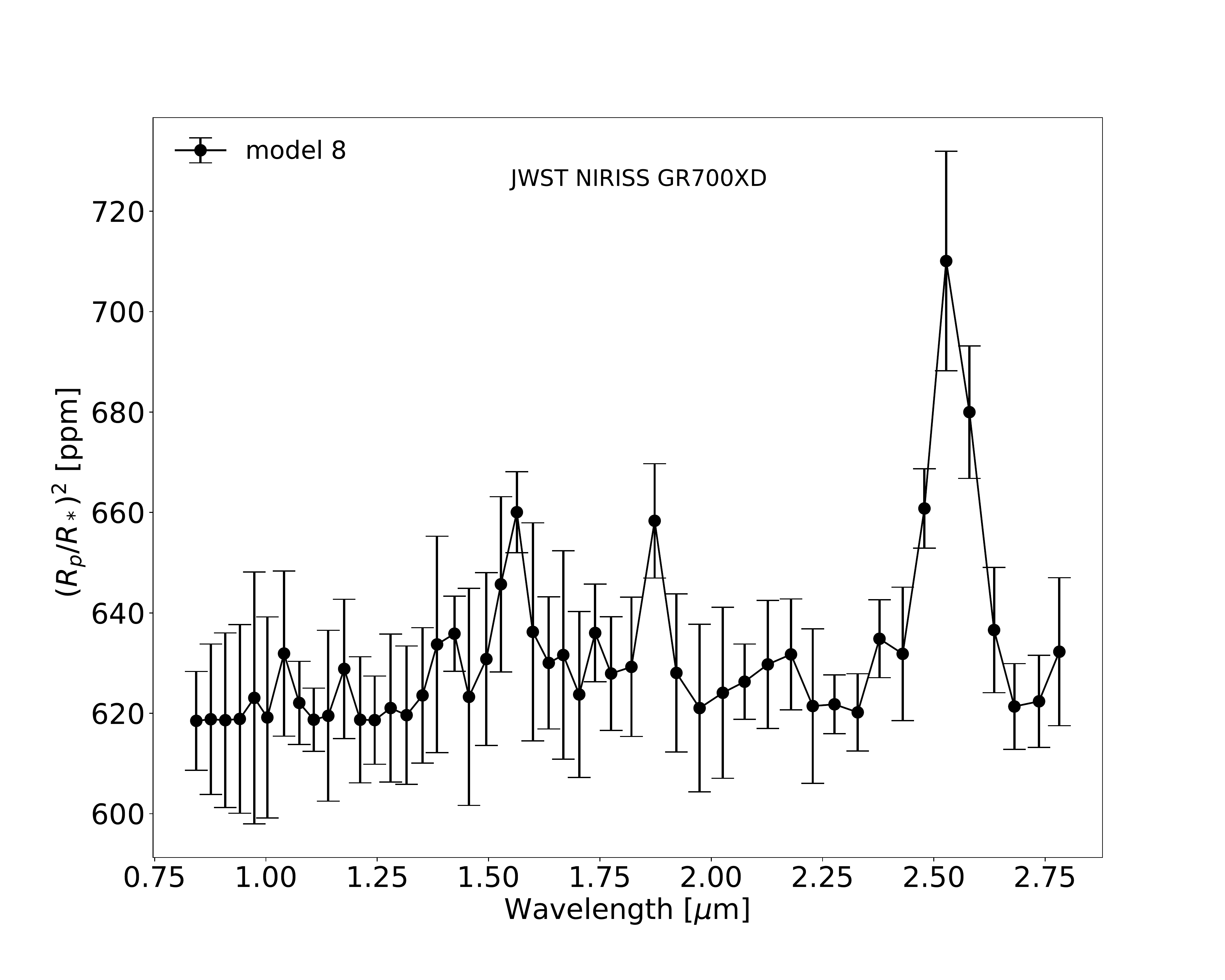}
\caption{Simulated spectrum of JWST/NIRISS GR700XD. Theoretic spectral model of L 98-59 b in the wavelength range of JWST/NIRISS GR700XD for a single transit visit. If HCN is present, JWST will be able to confirm its presence.}
\label{jwst}
\end{figure}

Future missions such as the James Webb Space Telescope (JWST) and Ariel \citep{tinetti18} will provide higher sensitivity and/or wider spectral coverage than HST/WFC3. This will have a huge impact on the detection of atmospheric features on small exoplanets, even with the presence of clouds at high altitudes \citep{Hinkley22, Constantinou22, Whittaker22}. 
If there is a significant amount of HCN in the atmosphere of L~98-59~b, we should be able to detect a very strong absorption feature near 2.5~$\mu m$, which is covered by the wavelength ranges of instruments from both of JWST and Ariel. 
Although we do not consider the model with abundant HCN in the atmosphere of L~98-59~b to be real, we still decide to utilize it as an example to demonstrate the power of JWST here.
The cloudy HCN+$N_2$ model (model 4) is used to simulate the spectrum from one single-transit observation of L~98-59~b using NIRISS GR700XD spectrograph (with a resolution of R~$\sim$ 50) on JWST. The simulated spectrum is presented in Figure~\ref{jwst}. Using the plot, we can demonstrate that observations from JWST can be used to place strong constraint on the HCN abundance in L~98-59~b.

\subsection{Stellar Contamination}

There is no evidence of strong stellar activity present on L~98-59 from the TESS data \citep{2019AJ....158...32K}, and it is likely to be a quiet M-dwarf with a weak level of XUV activity \citep{2021AJ....162..169P}. 
However, since very high precision measurements are needed for this study, we still need to quantify the potential contamination to the transmission spectrum of L~98-59~b from the stellar activity of L~98-59.

We apply the stellar contamination models from \cite{2018ApJ...853..122R} to calculate the potential effect of the stellar activity (including spots and faculae) on the transmission spectrum of L~98-59~b.
Here we take four cases of stellar activity into account: giant spots, solar-like spots, giant spots with faculae, and solar-like spots with faculae.
In \cite{2018ApJ...853..122R}, they define that the radii of giant spots and solar-like spots are $ R_{spot} = 7^\circ$ and $ R_{spot} = 2^\circ$, respectively.
We compute the effect of stellar contamination on the transmission spectrum at different wavelengths using Equation (3) from \cite{2018ApJ...853..122R}. For each case, the spot and faculae covering fractions for a M3 type star are adopted from \cite{2018ApJ...853..122R}. The temperatures of photosphere, spot, and faculae are assumed to be 3200 K, 2800 K, and 3300 K, respectively, with a surface gravity of log($g$) = 5.0 and metallicity of [Fe/H] = -0.5. We use the theoretical PHOENIX BT-Settl spectra for photosphere, spot, and faculae.
The final contaminated transmission spectrum is derived by multiplying the transit model of L~98-59~b. 

\begin{figure}
\gridline{\fig{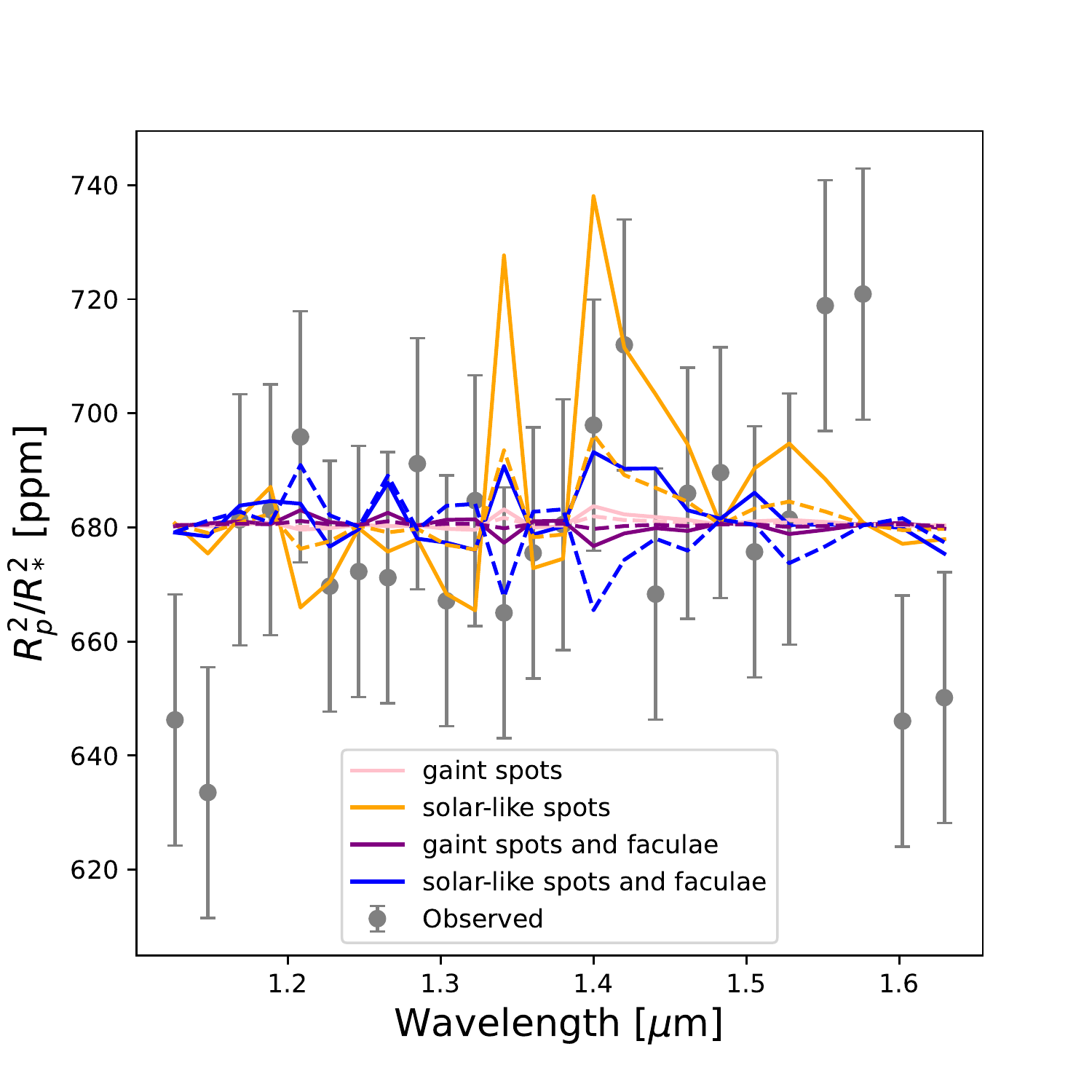}{0.5\textwidth}{}}
\caption{Contamination of the transmission spectrum of L~98-59~b from different stellar activity models. Lines with the same color represent stellar contamination models with different filling factors. 
The solid and dashed lines represent the transit light source effect for the maximum and mean spot or faculae filling factor,
respectively.}
\label{stellar contamination}
\end{figure}

To check which stellar contamination model can best explain the observed transmission spectrum, we compare the observed spectrum with different stellar contamination models in Figure~\ref{stellar contamination}. In this figure, solid and dashed lines represent the transit light source effect for the maximum and mean spot or faculae filling factor, respectively \citep{2018ApJ...853..122R}. 
We have also calculated the $\chi^2$ and reduced $\chi^2$ to assess the goodness of fitting for different models, which are summarized in Table~\ref{chi of stellar contaminations}. 
From Figure~\ref{stellar contamination} and Table~\ref{chi of stellar contaminations}, we can see that the model of solar-like spots with the maximum spot filling factor deviates most from the observed data. Other stellar models' reduced-$\chi^2$ is $\sim$ 1, indicating that the stellar models can adequately describe the observed spectrum.  

After removing the impact of stellar contamination, we also re-run the atmospheric retrieval using model 4 and model 5.
The retrieval results are shown in Table~\ref{stellar contaminations}. 
We can see that for all the stellar models, the retrieval results are similar to those done in previous sections.
For one particular case, with the solar-like spots model, the pure cloudy model yields the best fitting result.
Therefore, we reach similar conclusion after including the effect of stellar contamination in our retrieval analysis.

\begin{table}[t]
\centering
\caption{The $\chi^2$ and reduced-$\chi^2$ for different stellar contamination models. \label{chi of stellar contaminations}
 }
\begin{threeparttable}
\resizebox{0.5\textwidth}{!}{
\begin{tabular}{l|l|l|c}
\hline
\multicolumn{2}{c|}{Stellar model} & $\chi^2$ & reduced-$\chi^2$ \\
\hline
\multirow{2}{*}{Giant spots} & max & 22.887 & 1.040 \\
\cline{2-4}
& mean & 23.082 & 1.049 \\
\hline
\multirow{2}{*}{Solar-like spots} & max & 33.264 & 1.512 \\
\cline{2-4}
& mean & 22.504 & 1.023 \\
\hline
\multirow{2}{*}{\shortstack{Giant spots\\+ faculae}} & max & 23.705 & 1.129 \\
\cline{2-4}
& mean & 23.314 & 1.110 \\
\hline
\multirow{2}{*}{\shortstack{Solar-like spots\\+ faculae}} & max & 22.087 & 1.052 \\
\cline{2-4}
& mean & 26.141 & 1.245 \\
\hline
\hline
\end{tabular}}
\end{threeparttable}
\end{table}

\begin{table*}
\centering
\caption{The retrieval results assuming different stellar contamination models.}
\label{stellar contaminations}
\begin{tabular}{l|l|l|l|l|l|l}
\hline
\multicolumn{2}{c|}{\multirow{2}*{Stellar model}} & \multicolumn{1}{c|}{\multirow{2}*{HCN abundance}} & \multirow{2}{*}{log(E) of model 4} & \multirow{2}{*}{log(E) of model 5} & 
\multicolumn{1}{c|}{\multirow{2}{*}{\shortstack{$\Delta$ log(E)}}} & \multirow{2}{*}{\shortstack{Sigma}}\\
\multicolumn{1}{c}{} & \multicolumn{1}{c|}{} & &  &  & & \\

\hline
\multirow{2}{*}{Giant spots} & max & -0.63$^{+0.45}_{-3.42}$ & 229.799 $\pm$ 0.063 & 228.931 $\pm$ 0.055 & 0.868 & 1.940
\\
\cline{2-7}
& mean & -0.62$^{+0.45}_{-3.02}$ & 229.820 $\pm$ 0.064 & 228.700 $\pm$ 0.055 & 1.120 & 2.095\\
\hline
\multirow{2}{*}{Solar-like spots} & max & -4.01$^{+3.74}_{-5.17}$ & 228.586 $\pm$ 0.059 & 229.052 $\pm$ 0.055 & - & - \\
\cline{2-7}
& mean & -0.77$^{+0.55}_{-5.46}$ & 230.280 $\pm$ 0.061 & 229.893 $\pm$ 0.055 & 0.387 & 1.572 \\
\hline
\multirow{2}{*}{\shortstack{Giant spots\\+faculae}} & max &-0.63$^{+0.44}_{-1.96}$ & 229.672 $\pm$ 0.065 & 228.115 $\pm$ 0.055 & 1.557 & 2.331\\
\cline{2-7}
& mean & -0.57$^{+0.40}_{-2.56}$ & 229.770 $\pm$ 0.064 & 228.464 $\pm$ 0.055 & 1.306 & 2.120\\
\hline
\multirow{2}{*}{\shortstack{Solar-like spots \\+faculae}} & max & -0.70$^{+0.51}_{-3.83}$ & 230.814 $\pm$ 0.063 & 230.104 $\pm$ 0.055 & 0.71 & 1.833\\
\cline{2-7}
& mean & -0.58$^{+0.41}_{-0.68}$ & 228.999 $\pm$ 0.067 & 225.882 $\pm$ 0.056 & 3.117 & 2.993 \\
\hline
None & - & -0.63$^{+0.44}_{-2.51}$ & 229.740 $\pm$ 0.064 & 228.540 $\pm$ 0.055 & 1.2 & 2.14\\
\hline\hline
\end{tabular}
\end{table*}

\subsection{Primordial H/He Atmosphere and Mass Loss}
\label{sec:loss}

According to \citet{owen20}, low-mass planets similar to L~98-59~b can accrete about 1$\%$ of their total masses in H/He gases during formation.
L~98-59 is estimated to be $>$1~Gyr in age \citep{kostov19b}, which is inferred from the star's lack of spectroscopic youth indicators, slow rotation, HR diagram position, no evidence of spots, and low levels of white-light flare activity from TESS data.
According to the work of \citet{2021AJ....162..169P}, the relative size and insolation flux of L~98-59~b may have caused it to lose a significant amount of atmosphere. 
Therefore, assuming no other secondary sources of hydrogen, it is very unlikely that L~98-59~b could possess a present-day cloudy primordial H/He-rich atmosphere. However, data from WFC3 alone cannot be used to rule out this possibility entirely.

\section{Conclusion \label{sec:conclusion}}
We present the HST/WFC3 transmission spectrum analysis of the rocky planet L~98-59~b in this paper.
Using the Iraclis data reduction pipeline and the TauREx3 atmospheric retrieval code, we find the transmission spectrum can be fitted either using a flat-line pure cloudy model or a cloudy HCN atmospheric model.
It is very unlikely to possess a clear and hydrogen-dominated atmosphere. The existence of other volatile molecular species can not be confirmed because of the limited precision and wavelength coverage of the WFC3 observation. 
However, we cannot explain the unrealistically high HCN abundance derived using the equilibrium chemistry model. Therefore, we conclude that L~98-59~b is more likely to have a secondary thin atmosphere with a high  mean molecular weight, an atmosphere with a highly opaque aerosol layer, or effectively no atmosphere. These different scenarios cannot be distinguished using the current HST data. Future observation from JWST are needed to truly understand the existence and contents of an atmosphere around L~98-59~b.

\begin{acknowledgments}
We thank our anonymous referee for prompt and insightful comments which led to significant improvement of the manuscript. We thank professor Giovana Tinetti, whose visit has triggered this study. We thank Angelos Tsiaras, Ingo Waldmann and Ahmed Al-Refaie for their instructions on how to use Iraclis and Taurex. We acknowledge the financial support from the National Key R\&D Program of China (2020YFC2201400), China Postdoctoral Science Foundation (No.2020M672936), NSFC grant 12073092, 12103097, 12103098, 11733006, the science research grants from the China Manned Space Project (No. CMS-CSST-2021-B09), Guangdong Major Project of Basic and Applied Basic Research (Grant No. 2019B030302001), Guangzhou Basic and Applied Basic Research Program (202102080371), and the Fundamental Research Funds for the Central Universities, Sun Yat-sen University.

This work is based on observations with the NASA/ESA Hubble Space Telescope,  obtained at the Space Telescope Science Institute (STScI) operated by AURA, Inc. The publicly available HST observations presented here were taken as part of proposal 15856, led by Thomas Barclay. These were obtained
from the Hubble Archive, which is part of the Mikulski Archive
for Space Telescopes. This paper also includes data collected by the TESS mission,
which is funded by the NASA Explorer Program. TESS data is 
also publicly available via the Mikulski Archive for Space
Telescopes (MAST).
\end{acknowledgments}

	{	\small
		\bibliographystyle{apj}
		\bibliography{L98-59b.bib}
	}

\end{document}